\documentclass[showpacs,aps,prl,twocolumn,superscriptaddress]{revtex4} 
\usepackage[pdftex]{graphicx} 
\usepackage{amsmath} 
\usepackage{amsfonts} 
\usepackage{amssymb} 
\usepackage{amsbsy} 
\usepackage{ulem} 
\usepackage{color} 
\usepackage{times} 
\usepackage{units} 
\def\etal{$\it{et~al}$} 
\def\De{\Delta \text{E}} 
 
\def\dt{\Delta \text{T}/\text{T}} 
\def\nf{\nicefrac{1}{2}}

\begin{document} 

\title{Interaction-induced shift of the cyclotron resonance in graphene via infrared spectroscopy} 

\author{E. A. Henriksen} 
\altaffiliation{Electronic address: erikku@caltech.edu} 
\affiliation{Department of Physics, Columbia University, New York, New York 10027, USA}  

\author{P. Cadden-Zimansky} 
\affiliation{Department of Physics, Columbia University, New York, New York 10027, USA} 
\affiliation{National High Magnetic Field Laboratory, Tallahassee, Florida 32310, USA}  

\author{Z. Jiang} 
\affiliation{School of Physics, Georgia Institute of Technology, Atlanta, Georgia 30332, USA} 

\author{Z.Q. Li}
\affiliation{Department of Physics, Columbia University, New York, New York 10027, USA}

\author{L.-C. Tung} 
\affiliation{National High Magnetic Field Laboratory, Tallahassee, Florida 32310, USA}  

\author{M. E. Schwartz} 
\affiliation{Department of Physics, Columbia University, New York, New York 10027, USA} 

\author{M. Takita} 
\affiliation{Department of Physics, Columbia University, New York, New York 10027, USA}  

\author{Y.-J. Wang} 
\affiliation{National High Magnetic Field Laboratory, Tallahassee, Florida 32310, USA}  

\author{P. Kim} 
\affiliation{Department of Physics, Columbia University, New York, New York 10027, USA}  

\author{H. L. Stormer} 
\affiliation{Department of Physics, Columbia University, New York, New York 10027, USA} 
\affiliation{Department of Applied Physics and Applied Mathematics, Columbia University, New York, New York 10027, USA} 
\affiliation{Bell Labs, Alcatel-Lucent, Murray Hill, New Jersey 07974, USA}  

\date{\today}  

\begin{abstract}
We report a study of the cyclotron resonance (CR) transitions to and from the unusual $n=0$ Landau level (LL) in monolayer graphene.  Unexpectedly, we find the CR transition energy exhibits large (up to 10\%) and non-monotonic shifts as a function of the LL filling factor, with the energy being largest at half-filling of the $n=0$ level.  The magnitude of these shifts, and their magnetic field dependence, suggests that an interaction-enhanced energy gap opens in the $n=0$ level at high magnetic fields.  Such interaction effects normally have limited impact on the CR due to Kohn's theorem [W. Kohn, Phys. Rev. {\bf 123}, 1242 (1961)], which does not apply in graphene as a consequence of the underlying linear band structure.
\end{abstract}

\maketitle  

The intriguing electronic properties of graphene in a strong magnetic field were first demonstrated by the observation of an extraordinary `half-integer' quantum Hall effect \cite{novoselov:197, zhang:201, zheng:245420,gusynin:146801,peres:125411}, resulting from the existence of an unusual $n=0$ Landau level (LL).  Later evidence from transport measurements indicates that the four-fold degeneracy of this $n=0$ level is completely lifted in high magnetic fields \cite{zhang:136806,jiang:106802}, and a gapped state forms \cite{checkelsky:206801,giesbers:0948,li:176804,miller:924}.  This state and the origin of degeneracy-breaking in the $n=0$ level is a matter of intense theoretical study, with electron-electron interactions expected to play a critical role \cite{yang:27}.  However, since the charge transport in earlier experimental works is either local, or dominated by the sample edge, direct access to the intrinsic properties of graphene in the bulk remains limited.  Additionally, the diverging resistance near half-filling of the $n=0$ level prohibits investigation of gapped states by thermal activation experiments.  In this Letter, we report a study of the $n=0$ LL utilizing infrared (IR) magnetospectroscopy, which is sensitive to the cyclotron orbits of charge carriers throughout the entire graphene sheet. As the LL filling factor is changed, we observe unexpected and sizable shifts in the cyclotron resonance (CR) transition energies.  We interpret this as due to electron-electron interactions which create a gap in the $n=0$ LL, thereby affecting the energies of CR transitions to and from this level. 

In a perpendicular magnetic field, $B$, the LL structure of graphene has an unusual energy dependence on both $B$ and the LL index, $n$, described by \cite{mcclure:666,semenoff:2449,haldane:2015} \begin{equation} 
\text{E}_n = \text{sgn}(n)\sqrt{2|n|} \hbar \tilde{c} / l_B~. 
\end{equation} 
Here $\hbar$ is Planck's constant, $\tilde{c} \sim 1\times10^6$ m/s the band velocity, and $l_B=\sqrt{\hbar/(e B)}$ the magnetic length.  The index $n$ runs over both positive and negative integers representing electron and hole-like states, and includes a field-independent $n=0$ LL which is located at zero energy.  This level, in combination with a four-fold level degeneracy due to the two-fold electronic spin and pseudo-spin (valley) degeneracies, leads to a unique sequence of LL filling factors, $\nu=\pm2,\pm6....$.  Here $\nu=2 \pi n_s l_B^2$, where $n_s$ is the carrier density.  In contrast to conventional 2D systems, in which even the lowest LL experiences an energy shift with $B$, in graphene the $n=0$ LL remains put at zero energy, is shared by the conduction and valence bands and, at charge neutrality for vanishing $\nu$, is filled half with electrons and half with holes.  

\begin{figure}
\includegraphics[width=0.9\columnwidth]{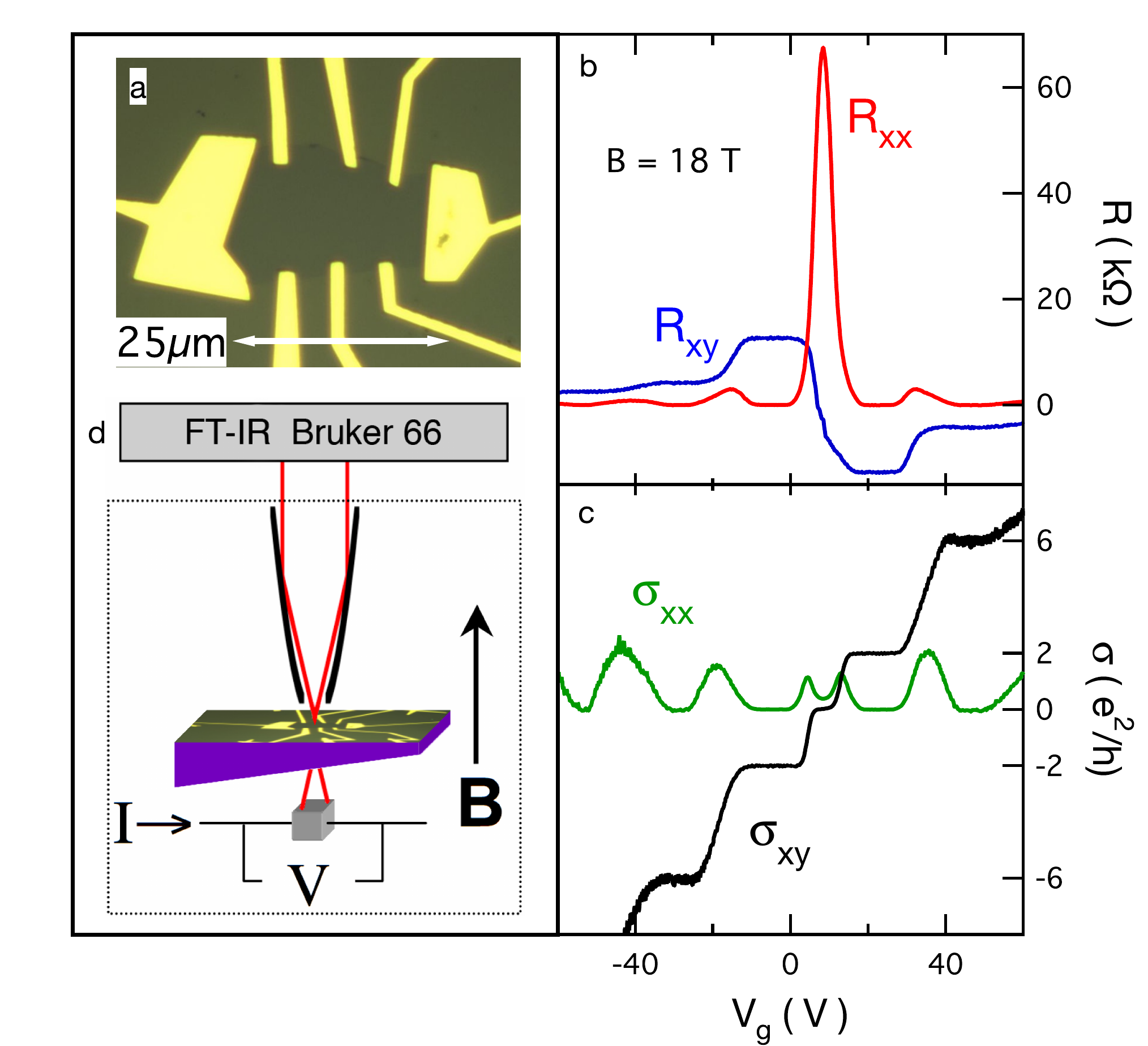} 
\caption{(a) Optical microscope image of sample A.  (b) Measured Hall (blue) and longitudinal (red) 4-wire resistances of sample A.  (c) Calculated Hall (black) and longitudinal (green) conductivities, showing a Hall plateau forming at $\nu=0$.  (d) Schematic of experimental setup showing IR light focused through graphene.  The varying bolometer resistance tracks changes in the transmitted light.} 
\end{figure}  

Motivated by recent experiments \cite{zhang:136806,jiang:106802,checkelsky:206801,giesbers:0948,li:176804,miller:924}, we have performed a careful study of LL transitions to and from the $n=0$ level via CR \cite{gusynin:157402,jiang:197403,sadowski:266405,deacon:081406,henriksen:087403,orlita:267601}.  The five samples, A-E, used in this work consist of monolayer graphene exfoliated from Kish graphite onto Si/SiO$_2$ wafer substrates \cite{novoselov:10451}, with Cr/Au contacts defined by standard thin film processing for {\it in~situ} measurement of sample transport properties.  Sample sizes and mobilities (at a hole density of 1.7$\times$10$^{12}$ cm$^{-2}$) range from 330-1,100 $\mu$m$^2$ and 3,000-17,000 cm$^2$/(Vs), respectively.  The lightly-doped Si substrates transmit mid-IR light, but remain sufficiently conductive for use as a back gate to vary the carrier density.  Samples are mounted at the focus of a parabolic mirror and cooled via He exchange gas in a probe inserted into a liquid He cryostat.  Broadband IR light from a Bruker IFS 66 Fourier transform spectrometer reaches the graphene via direct optics, and the transmitted light is detected by a composite Si bolometer placed immediately below the sample; see Fig. 1 (d).  The samples are wedged to suppress Fabry-Perot interference.  We determine the CR spectrum at constant $B$ field as in Ref.~\cite{jiang:197403}, recording IR transmission through the sample at two gate voltages, V$_\text{g}$ and V$_\text{b}$.  The change in transmission, $\Delta\text{T/T} = 1-\text{T(V)} / \text{T(V}_\text{b}\text{)}$, is measured over a range of V$_\text{g}$ ($\nu$).  The background scans, $\text{T(V}_\text{b}\text{)}$, are at $\nu=\pm10$.   

In Fig. 1 (a) we show an optical microscope image of our highest mobility monolayer graphene device, sample A, with mobility $\mu = 17,000$ cm$^2$/(Vs).  The Hall and longitudinal resistances for this sample, and the calculated conductivities, are shown as a function of V$_\text{g}$ in Fig. 1 (b) and (c) respectively.  A plateau in the Hall conductivity is clearly evident near $\nu=0$, as in Ref.~\cite{zhang:136806,checkelsky:206801,giesbers:0948}.  

In Fig. 2 (a) we plot the change in IR transmission, $\dt$, through sample A for a range of $\nu$ at a field of $B=18$ T.  From bottom to top, the filling factor decreases from $\nu=+4\nf$ to $\nu=-3\nf$ in steps of $\Delta \nu = -\nf$.  Physically, this amounts to shifting the Fermi level, E$_F$, through a fixed LL spectrum starting just above half-filling of the $n=1$ LL and ending below half-filling of the $n=-1$ LL.  The broad peaks correspond to the resonant excitation of carriers between the lowest LLs, via the $n=-1\to0$ and/or the $0\to1$ transition.  The sharper spikes at 157 meV and 196 meV are due to harmonics of 60 Hz electrical noise.  Additional small features resembling peaks or shoulders on the lineshape are visible in several traces. These are suggestive of splittings of the CR line.  Such an observation would be intriguing and underscores the need for higher quality samples, but here we focus on the main CR peak energies traced by the dashed line in Fig. 2 (a).

Closer inspection of the CR traces reveals a distinct asymmetry in the lineshape with the high-energy side falling off more slowly than the low-energy side. This distortion of the Lorentzian resonance originates in multiple reflections between the graphene and the wedged Si/SiO$_2$ substrate \cite{abstreiter:2480}.  To extract the physical parameters for each resonance, we have performed detailed curve fits by modeling the transmission through a multilayer system including substrate phonon modes and multiple reflections/refractions at the graphene-SiO$_2$ and SiO$_2$-Si interfaces.  The dielectric properties of the substrate used in the fits are determined experimentally by a combination of IR spectroscopy and ellipsometry \cite{li:532}.  Calculated fits are compared to the data in successive iterations, as the model resonance energy, intensity and width are varied until best fits are achieved as determined by inspection.  An example is shown in Fig. 2 (b) for the $\nu=+1\nf$ trace.  We note the peak energies from this fitting fall at energies slightly below the intensity maxima \cite{note}.

Far from remaining constant as the occupancy of the $n=0$ LL is changed, the CR exhibits a marked non-monotonic dependence on $\nu$, resulting in a nearly symmetric ``W''-shape as shown in Fig. 2 (c).  Two sharp minima in the CR energy occur at $\nu=\pm2$, with a more rounded maxima at $\nu=0$ that is nearly 10\% higher than the values at $|\nu|=2$.  We note the increase in energy occurring for $|\nu|>2$ is accompanied by a loss in intensity, with the peaks fading away beyond $|\nu|=4$.  This is due to the onset of the next LL transition, $n=1\to2$ or $-2\to-1$, as E$_F$ is shifted farther from charge neutrality and either the $n=+1$ or $-1$ LL becomes partially occupied. 

\begin{figure}
\includegraphics[width=\columnwidth]{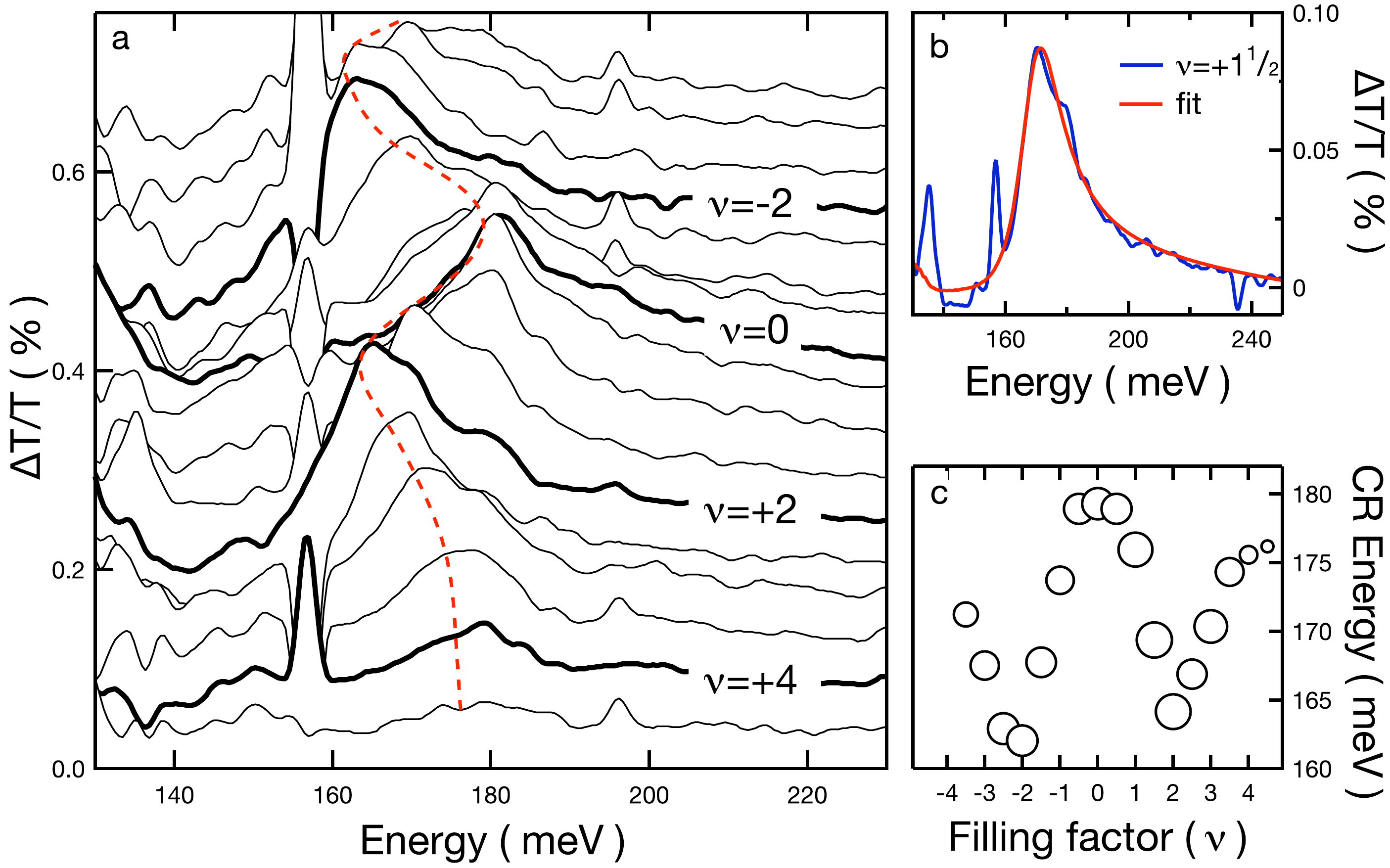} 
\caption{(a) The change in transmission, $\dt$, for CR traces in monolayer graphene at fixed field $B = 18$ T over a range of $\nu$.  Traces are offset for clarity.  The bottom trace is $\nu= 4\nf$; $\nu$ of successive traces decreases by $\nf$.  The dashed line is a guide to the eye, marking the peak energies determined by curve fits.  (b) Example of a fit (solid red line) to the CR line shape at $\nu=1\nf$ using a multilayer IR transmission model. (c) CR peak energies plotted vs. $\nu$.  The area of the markers represents the relative intensity.} 
\end{figure}  

The considerable shifts seen in Fig. 2 (c) are not reflected in the interband transitions that are excited simultaneously, albeit at higher energies.  In Fig. 3 we show the CR data for $\nu=\pm2$ and $0$ in sample B over an extended energy range which also includes the first degenerate pair of interband transitions, $n=-2\to1$ and $-1\to2$.  The transitions responsible for the two resonances in each trace are illustrated schematically in the upper left inset to Fig. 3.  The interband peak (at 440 meV) shifts by less than 1\% as $\nu$ is changed.  In contrast, and consistent with the behavior seen in Fig. 2, the lower energy peaks (at 165-180 meV) show the same behavior as in sample A, with a strong upshift ($\approx8$\%) in energy at $\nu=0$.

\begin{figure}
\includegraphics[width=\columnwidth]{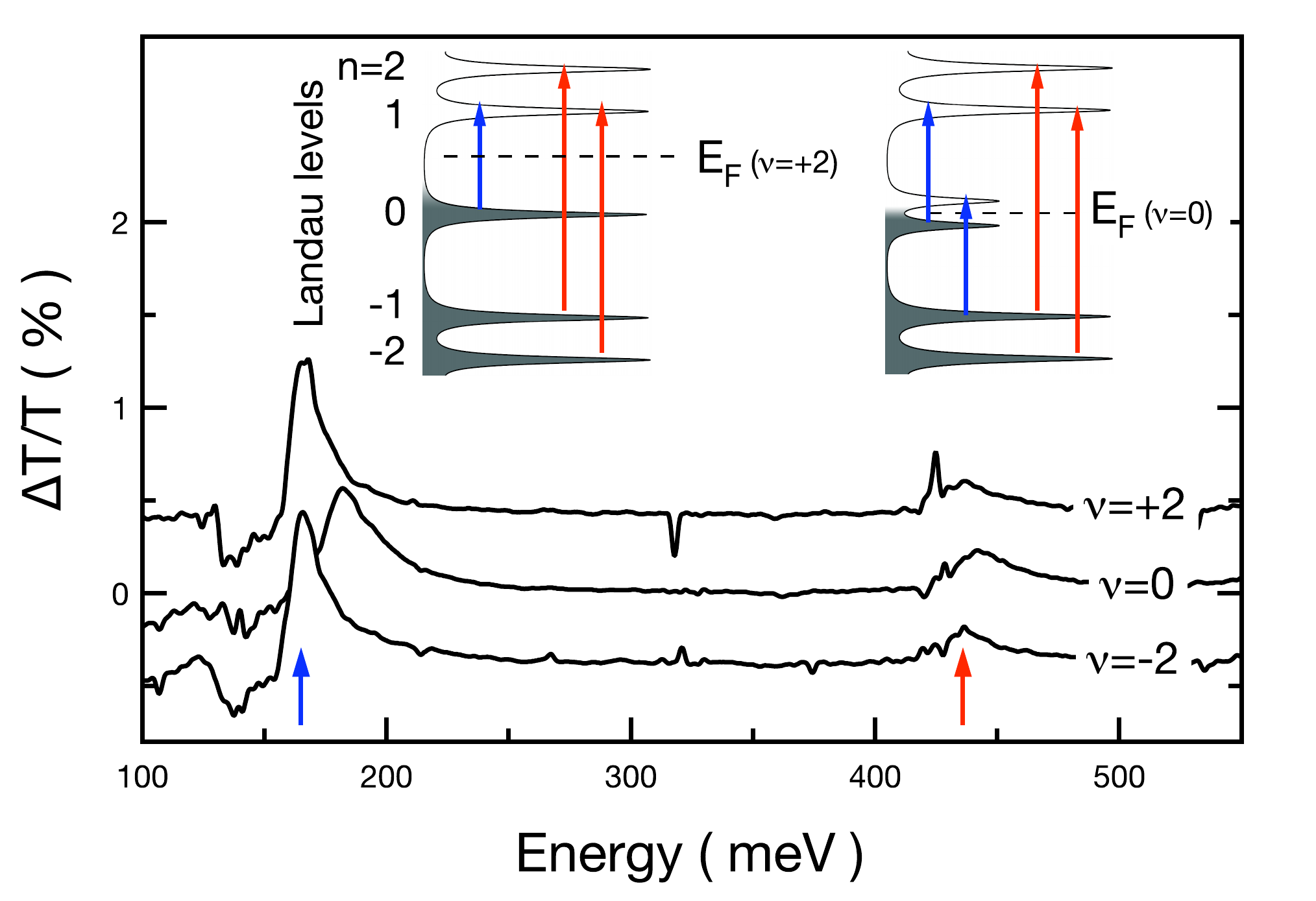} 
\caption{Two CR peaks visible in $\dt$ traces from sample B, at $B=18$ T.  The underlying transitions are $n=0\to1$ and/or $n=-1\to0$ near 170 meV (blue arrows), and $n=-1\to+2$ and $n=-2\to+1$ at 440 meV (red arrows).  Traces are offset for clarity.  Left inset: LL schematic showing transitions at $\nu=+2$, when E$_F$ lies between the zeroth and first LLs.  Right inset: proposed LL schematic at charge neutrality ($\nu=0$), showing a gap in the $n=0$ LL and the corresponding lowest energy CR transitions.}
\end{figure}  

The increase in CR energy of the lowest LL transitions at $\nu=0$ depends on both $B$ and the sample quality.  Figure 4 shows the $B$ field dependence in all five samples at fields up to 31 T.   Here we characterize the overall magnitude of the shift, $\Delta E$, as the difference in CR energies at $\nu=+2$ and $\nu=0$.  Open symbols show data from the highest mobility sample, A, and filled symbols are data from samples B-E.  A clear trend of increasing $\Delta E$ (up to 23 meV) with increasing $B$ emerges.  We find $\Delta E$ is correlated with device quality: sample A shows the largest shifts, while lower mobility samples have consistently smaller $\Delta E$.  Although the broader resonances in the lower mobility samples lead to wide error bars, in sample A the uncertainty is sufficiently small to explore a functional dependence between $\Delta E$ and $B$.  In the inset to Fig. 4, we plot the CR energies of the lowest LL transitions vs. $\sqrt{B}$ at two fields for sample A.  As suggested by Eq. 1, these data are well described by straight lines constrained to pass through zero and differing only in slope, implying that a $\sqrt{B}$ dependence is followed in both cases.  Further experiments in higher quality samples are needed to determine the precise field dependence of $\Delta E$.

\begin{figure}[t] 
\includegraphics[width=0.9\columnwidth]{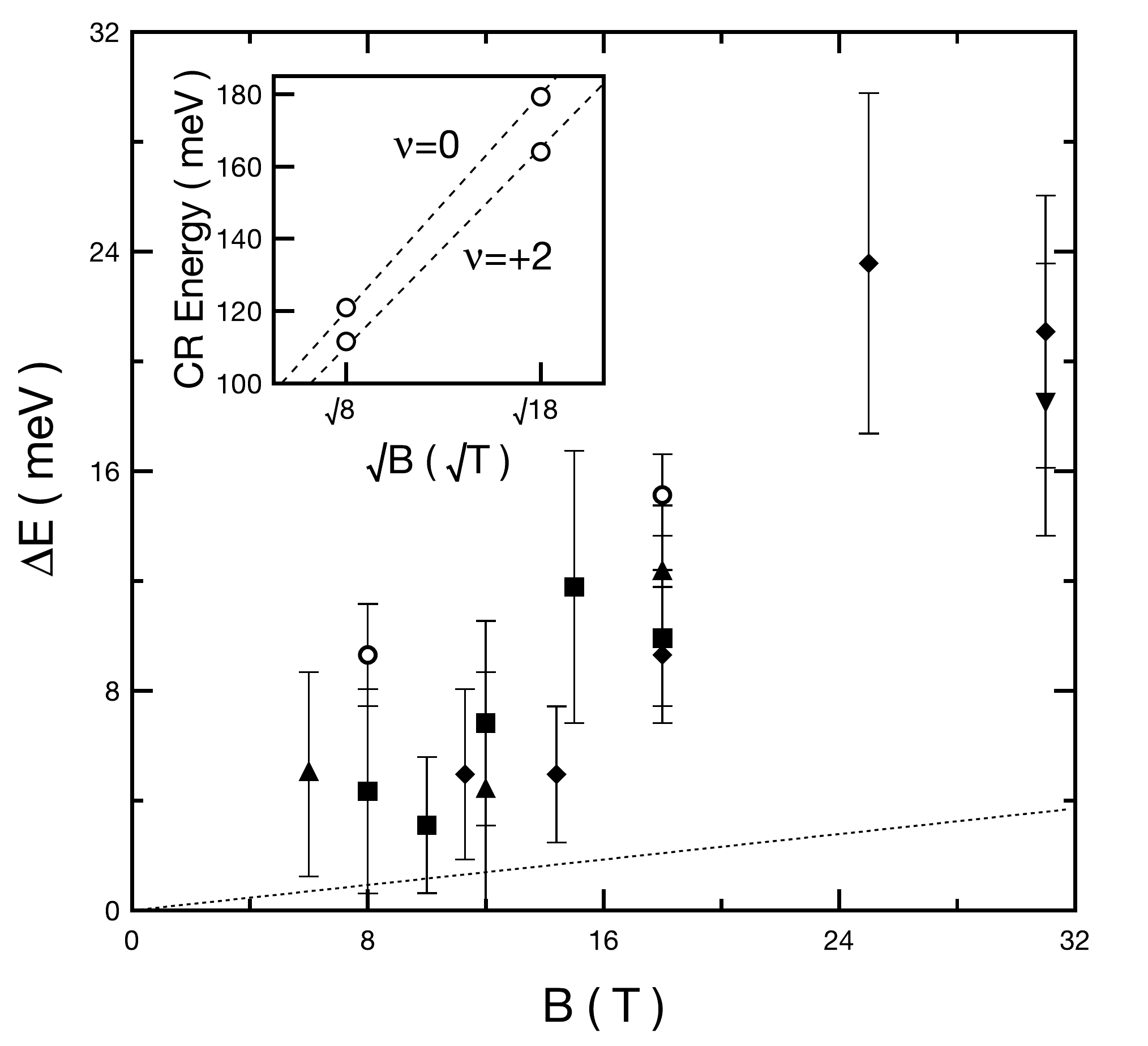} 
\caption{$B$ field dependence of the CR shift, $\De$, between $\nu=+2$ and $0$, as a function of $B$ for five samples.  Open symbols are from the high mobility ($\mu = 17,000$ cm$^2$/(Vs)) sample A resonances shown in Fig. 2; filled symbols show lower mobility samples ($\mu \leq 10,000$ cm$^2$/(Vs)).  The dotted line is the bare Zeeman splitting, $2 \mu_B B$.  Error bars are estimated by varying the best fit parameters.  Inset: $\nu=0$ and $+2$ CR energies of sample A at two fields, plotted against $\sqrt{B}$.  Dashed lines are least-squares linear fits through zero.} 
\end{figure}  

We now discuss possible origins of the observed CR shift at $\nu=0$.  In the basic model of non-interacting carriers, the CR energy is the difference of the initial and final LL energies from Eq. 1.  For $|\nu|<2$, two degenerate transitions occur, $n=-1 \to 0$ and $0\to1$.  For $2<\nu<6$ the$n=1\to2$ also occurs, but with an energy reduced by a factor of $\sqrt{2}+1$ that lies outside our experimental window.  Thus we expect to see only a single CR peak with energy independent of $\nu$, which clearly is not observed. 

Disorder certainly plays a role, as seen in Fig. 4.  We note a simple model where LLs are broadened only by short-ranged scattering predicts an upshift in the CR for half-filled LLs \cite{ando:989}, similar to our observation.  However, one would expect such a CR shift to increase in more disordered samples, in contrast to the observed increase in $\De$ with {\it increasing} mobility.  Also, the disorder potential in our samples is likely dominated by charged impurities \cite{chen:377}.  Screening may affect the CR transitions as well.  However, we expect increased screening to soften the CR \cite{schlesinger:73}, making for an M-shaped $\nu$-dependence rather than the W-shape of Fig. 2 (c); or else to narrow the resonance linewidth in half-filled levels \cite{heitmann:7463} which runs counter to the bulk of our data.  In this context, we note the broadening may be due to the linear dispersion alone \cite{mikhailov:241309}.

Single particle spin-related effects are also an unlikely origin for our observations.  Even at 31 T the bare Zeeman energy amounts to only 3.6 meV, several times smaller than the observed shifts (see Fig. 4).  Also, as spin is conserved in the CR transition, simple Zeeman effects should not be observed. There remains the possibility of a $\nu$-dependent g-factor enhancement. Large enhancements have been seen in other 2D systems \cite{nicholas:911}. However, these appear only in transport data with the interpretation invoking many-body effects.

The linear dispersion of graphene implies the CR can be influenced by many body effects \cite{bychkov:125417}.  In our experiments, the clear dependence of the CR energies on $\nu$-- and hence E$_F$-- and the magnitude of the shifts suggest that interactions at the Fermi surface are an important ingredient.  The contribution of electron-electron interactions to the dispersion of low-lying inter-LL excitations has been calculated for graphene in Ref.~\cite{iyengar:125430}, and also in Ref.~\cite{bychkov:125417} where the focus is restricted to those modes that contribute to CR.  In these works, interactions are generally found to increase the excitation energies.  In particular, the calculations show that when $\nu>2$ ($\nu<-2$), the $n=0\to1$ ($n=-1\to0$) transition develops a splitting where the higher energy peak continues to increase in energy with increasing (decreasing) $\nu$.  Qualitatively, this can describe the CR shifts we observe at $|\nu|>2$, if we assume the splitting is masked in our data by the broadened lineshape.  On the other hand, neither work predicts a change in the energy of optically active modes for $|\nu|<2$, and so fail to capture the large upward shift observed in our CR data near $\nu=0$.  Yet in spite of this, interactions seem to be the origin for our observations: although at 18 T the bare Coulomb energy $e^2/(\epsilon l_B) \approx 90$ meV is much larger than the shifts we observe in Fig. 2, no other energy scale is available.

Zhang \etal~ have observed quantum Hall plateaus at $\nu=0$ and $\pm1$ which they interpret as interaction-induced splittings of the $n=0$ level \cite{zhang:136806}. Naively, one could view our results as the consequence of such a splitting, because when the Fermi level passes through the $n=0$ LL, the presence of a gap leads to shifts in the transition energies to neighboring LLs; see the right inset to Fig. 3.  This picture, while certainly an oversimplification, is nonetheless appealing: it easily accounts for the $\nu$- and $B$-dependence of the CR energies, and is consistent with the fact that only those transitions involving the $n=0$ LL are strongly affected.  

As to similar experiments, the CR data on epitaxial graphene \cite{orlita:267601} have not yielded similar behavior.  Previous experiments on GaAs and Si/SiO$_2$ 2D systems have shown unexpected shifts and splittings of the CR bearing some resemblance to our data \cite{batke:3093,wilson:479,cheng:3171}.  However in detail the comparisons break down.  An increased effective mass of the carriers (decreased CR energy) seen for full LLs in 2D GaAs was attributed to a particular impurity doping, to band non-parabolicity, or to $\nu$-dependent screening, all of which are unlikely to be the source of our observations in graphene.  The origin of the anomalous CR peaks and splittings in the Si data \cite{cheng:3171} was never resolved, but in general these features had no $B$ dependence, with higher quality samples showing weaker anomalies, both in contrast to our data.

In conclusion, we have observed sizable shifts of the CR energy as a function of the LL filling factor and applied $B$ field for transitions involving the $n=0$ level in monolayer graphene.  We suggest our results indicate the opening of an interaction-induced gap in the bulk of the graphene at $\nu=0$ which increases with increasing $B$ field strength, and is detectable via CR as Kohn's theorem does not apply in graphene.  

We wish to thank M. Koshino, Kun Yang, Y. Barlas, A. MacDonald, S. Das Sarma, M. Fogler, M.-Y. Chou and J. Checkelsky for helpful discussions.  This work is supported by the DOE (DE-AIO2-04ER46133, DE-FG02-05ER46215 and DE-FG02-07ER46451), the NSF under DMR-03-52738 and CHE-0117752, ONR (N000150610138), NYSTAR, the Keck Foundation, the Microsoft Project-Q, and the SRC-NRI-MIND.  The IR work was performed at the National High Magnetic Field Laboratory, which is supported by NSF Cooperative Agreement No. DMR-0654118, by the State of Florida, and by the DOE.  We thank J. Jaroszynski, S. Hannahs, J.-H. Park, and E.C. Palm for experimental assistance.


\begin{thebibliography}{text}
\bibitem{note}The CR energies at $\nu=\pm2$ fall slightly below those from a lower mobility device \cite{jiang:197403}; however, the ratio of the energies of the two transitions visible in Fig. 3 remains greater than $1+\sqrt{2}$, as previously observed.
\end{thebibliography}
\end{document}